# A Review of Spatial Network Insights and Methods in the Context of Planning: Applications, Challenges, and Opportunities

Xiaofan Liang, Georgia Institute of Technology, xiaofan.l@gatech.edu
Yuhao Kang, University of Wisconsin – Madison

**Abstract** With the rise of geospatial big data, new narratives of cities based on spatial networks and flows have replaced the traditional focus on locations. While plenty of research that have empirically analyzed network structures, there lacks a state-of-the-art synthesis of applicable insights and methods of spatial networks in the planning context. In this chapter, we reviewed the theories, concepts, methods, and applications of spatial network analysis in cities and their insights for planners from four areas of concerns: spatial structures, urban infrastructure optimizations, indications of economic wealth, social capital, and residential mobility, and public health control (especially COVID-19). We also outlined four challenges that planners face when taking the planning knowledge from spatial networks to actions: data openness and privacy, linkage to direct policy implications, lack of civic engagement, and the difficulty to visualize and integrate with GIS. Finally, we envisioned how spatial networks can be integrated into a collaborative planning framework.

## 1 Introduction

The world we live in today is increasingly connected in physical, social, and economic ties. Our society has evolved into a "Network Society", a characterization from Castell [1], as being more decentralized, open, and organized as "a space of flows". These connections, which often congregate in large cities, are the keys to unlock economic agglomeration, innovations, and productivity beyond population growth and limited resources [2]. Therefore, urban planning as a discipline that makes future-oriented decisions for cities should grapple with this new social structure and its implications on human activities and urban forms.

_______________________

Yet, planning theories and tools to address cities as systems of network and flows have just started to be integrated at the cross of the century [3, 4]. This perspective of cities come from complexity science theory, which has been embraced in both collaborative (and communicative) planning theories [5, 6, 7] and a positivist approach to "the science of cities" [4, 2]. Collaborative planning theorists treat networks as the mechanism to empower the stakeholders to build consensus [8], while rationalists [4] see networks as the mechanism to model and explain urban growth and morphology. Tools that support this vision of cities come from a wide range of disciplines (e.g., agent-based models, social network analysis), most of which rely on individuals to interact and effect changes on collective phenomenon. Though critics argue that the technoscientific urbanism thinking embedded in the analytical and modeling approaches over-emphasizes technology's role at solving urban problems [9, 10], attentions to networks and flows as the background knowledge for planning keep rising, especially for planning smart cities [11]. Metrics derived from network measures such as vulnerability, reliability, and accessibility have also become the new normative goals for planning [12]. Under this context, spatial network analysis can be both interpreted as a tool to materialize a vision of city based on connectivity and a planning process that respects organic organizations and emergence from interconnected individuals and infrastructure.

Spatial networks (and network analysis) is a "fuzzy concept" [13, p.223] that lacks clear disciplinary boundaries. Its development has diverging roots in the field of network science (i.e., graph theory), social science (i.e., social network analysis), and geography (i.e., relational geography). Studies from network science perspectives tend to elaborate on the mathematical and formal elements of networks [22]. Social network analysis (SNA), originated from Sociology, has a long tradition focusing on how network structures inform power dynamics, group identity, and social relations [15]. While network science and SNA may not explicitly engage with agents' spatial locations, relational geography theories affirm that human relationships are spatially embedded in a global system of economic activities and institutional practices [16, 17]. Agents can reap the benefits of innovation, knowledge diffusion, and collaborations across geography or exert influence at a distance through their connections [18]. However, these virtual interactions are still not free from the constraints of geography [19].

Here, we define a spatial network as a graph structure in which nodes can be geolocated [20]. Spatial networks have two common types —— planar and nonplanar networks —— and they are differentiated by edges' spatial attributes. In a planar network, both vertices and edges are geographically embedded [20]. No edges will cross each other, and each intersection is a node, such as those in the road networks and electrical grid networks. A non-planar network has edges that overlap with each other without creating new nodes at the crossing [20]. These edges can represent non-spatial connections, such as social media friendship and telecommunications. Some networks can stand in between the two typologies, as their edges are somewhat constrained by geography but not completely embedded, such as flight routes and mobility patterns [20]. The separation of planar and nonplanar networks has implications on network statistics (e.g., degree distribution), as outlined in Haggett and Chorley's early classic *Network Analysis in Geography* [21] and later summarized in [22].

In this chapter, we contributed to the existing literature by associating and contextualizing the insights from spatial networks literature to planning concerns and described the challenges and opportunities to integrate spatial network analysis into planning practices. The following sections are divided into three parts. First, we divided the literature based on themes and reviewed the methods and insights from empirical studies (with a focus on the 2010-2020 period) that applied spatial network analysis to study urban dynamics. Second, we outlined four challenges of spatial network analysis revealed by the current work along with potential solutions. We concluded the paper with a vision framework for spatial networks to empower collaborative planning and examples to integrate spatial networks for different urban stakeholders. We focused our reviews on spatial networks that engage human activities.

## 2        A Review of Spatial Networks Literature for Planning Knowledge

The literature selected in the review represent a major line of research theme in spatial networks and was applied to increase planning knowledge. We consolidated the contents into four subsections based on their research goals.

### 2.1  Revealing Spatial Structures

Spatial network analysis has been used to reveal inter-city and intra-city spatial structures. The word "structure" is often referenced vaguely in the literature. Therefore, we divided the interpretations of structure into three kinds —— relationship, hierarchy, and cluster —— based on the research questions and the methods in the literature we reviewed. Many touched on more than one interpretation in the analysis.

Literature focusing on revealing the relational structure tend to ask how social and economic relationships and processes intersect with space. Thus, methods applied in this line of work rely on GIS and network visualizations heavily to contrast the spatial and social patterns or construct statistical indices to measure the effect of interests. For example, Graif [23] explained crime through neighborhood effect, a network of neighborhoods connected through gang members' daily commutes and social ties, rather than isolated incidents. People also form their own "gang turf" by consistently visiting certain places in cities for consumption or activities. Geographers and sociologists also refer to this spatial pattern prescribed by people's mobility traces and social media check-in data as the activity space. People or neighborhoods with various demographic labels can have distinct activity space patterns that reinforce social [24] and spatial segregation [25, 26]. The extent of segregation can be measured through the reciprocity and directions of mobility flows [27] or the overlap of demographic attributes [28].

When connections in spatial networks indicate more than just relationships, the network nodes start to embody influence and power. As a result, hierarchies arise as some nodes become more "central" or "influential". These characteristics can be measured through the distribution of network metrics (e.g., centrality) and the roles of nodes (e.g., hubs and spoke) at different levels. The three most widely applied centrality measures are degree centrality, closeness centrality, and betweenness centrality [29]. High degree centrality implies influence and vibrancy, as the node attracts a large amount of connections. High closeness centrality indicates accessibility, as the node can reach all other nodes in a few hops. High betweenness centrality represents low resilience, as the traversal of shortest paths from other nodes has to pass through a few nodes. These indices have been used as benchmarks to compare and group cities with similar characteristics in the urban context. For example, Taylor [30] 's world city network literature ranked cities by the number of offices from global firms (weighted by offices' importance) and found New York and London as the alpha cities in the world. Crucitti [31], on the other hand, used the centrality measures of urban street networks to distinguish self-organized and planned cities. Transportation literature also leveraged these indicators to assess urban infrastructure and predict traffic flows. For instance, Guimera [32] observed scale-free small-world characteristics in the flight network through degree distribution of airports, which means that people can reach any airport within a few layovers. Derrible [33] measured betweenness centrality of twenty-eight metro systems and found that the system becomes more resilient (i.e., more distributed betweenness centrality) when the number of stations increases. Though betweenness centrality is a prominent measure to predict traffic flows, its efficacy is debatable [34]. Also, a node with high centrality in one measure may not be high in another, which gives different roles to the nodes [35, 36]. Other than centrality measures, various network typologies and statistics are devised to capture the core-periphery structure in spatial networks, such as hubs and spokes [37], single-allocation networks [38] (or single-linkage analysis [39]), and rich-hub coefficients [40, 41].

When hierarchy is established, a natural next step is to cluster cities (or other nodes) to find communities in the networks. Questions asked in this type of work concern classifying various spatial units and visualizing enclaves that may not be spatially contiguous but deeply connected. Fortunato [42] summarized a series of community detection algorithms that can be applied to delineate areas in a network into fine-grained, non-overlapping regions. The key idea behind community detection is to separate nodes into distinct groups to minimize the within-group difference. One class of algorithms relies on traditional graph partitioning methods to either divisively remove edges that bridge between groups (e.g., Louvain method) or additively combine groups with similar characteristics (e.g., hierarchical clustering) [42]. The other class is based on optimizing modularity (i.e., quality of network partition), such as Fastgreedy [43], Spinglass [44], Walktrap [45], and Infomap [46]. Each method has its own strength given network size, edge directions, running speed, and efficacy [47]. These methods can be implemented in R and Python packages or network software like Gephi. In terms of applications, Ratti [48] adopted a spectral clustering algorithm (a way to optimize modularity) to "redraw" the boundaries of Great Britain based on cellphone interaction network data. Follow-up studies expanded the application to France, UK, Italy, Belgium, Portugal, Saudi Arabia, Ivory Coast, China, and Singapore [54, 55, 56]. The same approach was also applied on the neighborhood level to reveal intra-city dynamics [52, 53]. The generated borders tend to conform with administrative boundaries on the regional and national level, while less on the neighborhood level, with emerging urban centers featured by people's telecommunications and daily commutes [54, 55, 56]. One limitation of the modularity-based community detection method is that it only clusters based on the edges' origins and destinations and thus can be subjective to the modifiable areal unit problem (MAUP) and edge effects. More recent work uses linear units, such as GPS trajectories and streets, as the new focus for clustering [57, 58]. Boeing [59]'s OSMnx package significantly lowered the difficulty to analyze street networks by automating the data download, processing, visualization, and analysis with OpenStreetMaps. An example of the applications on the urban spatial order used this package to examine street networks' orientation, configuration, and entropy to group and compare cities with different urban forms [60].

## 2.2  Optimization of Urban Infrastructure

We wanted to feature a particular line of application that focused on planar networks and optimization methods. The research questions relevant to planning revolve around network traversals (e.g., most efficient network path to cover the problem area) and topology change (e.g., improve traffic and human flows with minimum road network changes).

   In transit planning, the network design of public transit is often framed as a bi-level optimization problem. Planners who wish to build the most efficient transit networks (first level optimization) must first resolve the users' demand to travel most efficiently (second level optimization). Spatial networks of human mobility can inform planners of user demands for transit and the current traffic bottlenecks. The transit routes can also be framed as spatial networks to be optimized for structural efficiency. For example, with GPS embedded into the bicycles and biking docks, the O-D flows, and user trajectories were collected to indicate demands and connectivity at various stops and thus optimize the locations of biking docks and construction of bike lanes [61, 62, 63, 64]. Other applications include reducing bus stop redundancy [65], evaluating operation efficiency in bus network [66], and designing spatial allocations of logistic centers [67]. The application of spatial networks on topology change is still developing. Brelsford [68] demonstrated how to generate alternative street network topology, which may improve urban slums' conditions.

## 2.3  Indicators of Economic Wealth, Social Capital, and Residential Mobility

Studies about the spatial structure or urban infrastructure tend to treat spatial network edges as homogeneous and assume they come from a single source. In fact, density, the types of edges, and the attributes of those they connected to all impact individuals or neighborhoods' social and economic health.

The foundation piece by Eagle [69] was the first of its kind to study the economic impacts of the network structure empirically. His finding indicated that neighborhoods' diversity of communication connections is strongly associated with their economic development. Evidence from highway transportation networks [70] also confirmed this positive association. Spatial networks can also inform an individual's (or a neighborhood's) levels of social capital based on the possession of far-reaching or local ties. For instance, Facebook friendship data told us that counties with fewer percentage of friends within 100 miles (e.g., San Francisco) are more likely to have higher social capital, social mobility, average income, and education levels [71]. On the individual level, these social supports can be maintained in sparse and transitive networks and is unaffected by residential moves [72], though the effect may be mediated by income or race. On the neighborhood level, social capital can be measured by the diversity and serendipity of visits, which formed the social diversity index formula in Hristova [73]'s paper.

The entanglement between race, income, and interpersonal network can also influence people's residential mobility and neighborhood mobilizations [26]. Conversely, the opportunity to grow and maintain a social network is partially contingent on the locations of the individual and ties. As Van Eijk [74] found out, poor people were more likely to have local ties. Living in a mixed (income or race) neighborhood did not necessarily bring rich connections due to the lack of interactions with resourceful neighbors. They were also more likely to form close-knit, kin-based social networks with geographic proximity, inhibiting information transmission and mobilizations for changes. Therefore, the destruction of low-income living communities is both "convenient" due to the lack of organized resistance, as evident in stories of 1960s urban renewals [75], and detrimental as it tears down locally-maintained social capital. The lack of ties outside of the poor's living communities may also explain why they were the last to evacuate (or not evacuate) their homes from natural disasters because of the difficulty to relocate [76].

## 2.4 Public Health Control

The Healthy Cities Movement started 40 years ago has grown planners' attention to managing public health in urban space. The outbreak of the COVID-19 global pandemic has posited a unique context for applying spatial network analysis for public health control and measuring the effect of policies that constrains human mobility.

The spatial structures and urban hierarchies embedded in spatial networks generate natural pathways for the contact-based disease to spread. Performing spatial network analysis helps governments and urban planners evaluate populations at risks, capture epidemiology-relevant behaviors, measure the effects of policies, and assist decision-making such as emergency response, medical resources evaluation and allocation, etc [77, 78]. Research has postulated that highly connected cities may be the first to be infected. Thus, spatial networks can be useful at tracking disease transmission and expecting populations at risk, as found in road [79], migration [80], and travel networks [81]. Several human mobility portals and datasets were created to support the spatial network visualization of the transmission pathways and human response to policies [82, 83, 84, 101]. Researchers have investigated how different lockdown strategies and intervention scenarios (e.g., social distancing) affect the spread of the disease and economic conditions by constructing human mobility flow networks in various countries including China [86], Italy [87, 88], France [89], UK [90], and U.S. [91].

The availability of spatial networks (e.g., mobility) data on the local level also enabled place-based epidemic modeling. Models that integrated spatial network structure and data (e.g., business foot traffic) were more successful at explaining the spatial heterogeneity in epidemic transmission across different regions and neighborhoods [92, 93, 94]. All these studies demonstrate the potential for spatial network structure to understand human behaviors in response to the disease and suggest the importance of incorporating spatial networks to inform health care planning.

In addition to policy evaluation, the COVID-19 public health crisis also recontextualized the flow characteristics associated with POIs (Point of Interests). As suggested by existing studies, mining POI characteristics and urban functions, such as the density of visits, the socioeconomic diversity of customers, and the types of interactions in place, are essential for urban planning [95, 96, 97, 98]. In the context of COVID-19, these characteristics are re-interpreted as transmission risks. For example, Benzell [99] incorporated the number of visits (and unique visits), time spent, and median distance traveled to different POIs to calculate transmission risk and ration what kind of places should be closed or reopen first. Knowing how POI characteristics may impact public health control, urban designers in the future may look into design solutions that address this challenge in advance.

## 3  Challenges and Opportunities

The application of spatial network analysis in the planning context did not come without challenges. Here, we discussed four constraints of spatial network analysis that need to be addressed for its further adoption in planning routines.

### 3.1  Data Openness and Privacy

Spatial network data are not widely accessible and consistently documented. Generally speaking, there are three commonly used data sources: private sources, crowdsourcing data, and the government released official records. Out of the fifty-seven empirical studies we reviewed in section 2, thirty-three (57%) used data from private sources, such as cell phone calls, location-based social network service, micromobility traces, GPS trajectories, and social media check-in. These data have the advantages to be fine-grained and have relatively high data quality but are often bought one-off for research purposes. They were disproportionally applied to study non-planar networks, such as social connections and mobility patterns (esp. in COVID-19). Six papers (11%) collected data through crowdsourcing, including volunteered GPS trajectories, open-sourced street networks, surveys, or collaboration with public institutions. These data tend to be small and costly to collect but more informative for specific research questions. OpenStreetMap (OSM) project as a well-known crowdsourced geographic data platform provides detailed road networks across the world, though the data quality may vary by region. Regarding government released official data sources, eighteen (32%) used publicly available data such as tax records, LODES commutes data from U.S. census, court records, street networks, bus routes, and smart cards. The attainment of some public data (e.g., air and train schedule) can involve time-consuming data scraping and cleaning. These public data also skewed heavily toward planar networks (eight papers) and have very little documentation for non-planar (social) networks. In addition, these data are often offered in aggregated formats, which is not fine-grained enough for planners' place-based work. The stark contrast of numbers above reveals the shortage of crowdsourced and publicly available spatial network data, especially for mobility and relationships. Considered the wide range of applications on urban affairs, spatial network data should be considered a public good despite being collected through private channels. So, how can we encourage data openness to increase the accessibility of spatial network data?

The City of Chicago provides an example of data openness through policymaking. Starting in 2018, all ride-share companies are required by an ordinance to send routine reports to the City of Chicago, including the origins and destinations of trips [100]. Products like Uber Movements also help transportation planners to monitor traffic flow and increase road safety. During the COVID-19 outbreaks, multiple geospatial data companies (e.g., SafeGraph) also contributed free and open POI or tract-based foot traffic data for governments, non-profits, and researchers to download [101]. Still, more public-private partnership models or policies need to be developed to keep spatial network data open.

Privacy, in particular, geoprivacy, is another barrier to collect large-scale and consistent spatial network data. This challenge is often offloaded to service vendors to resolve. Geoprivacy refers to an individual's rights to prevent the disclosure of sensitive personal locations such as their home, workplace, travel trips [102]. Due to the rapid development of location-based services, information, such as users' location records and attributes, is automatically collected or inferred from the spatial, temporal, and thematic characteristics of the geographic information. Hence, it is necessary to "encrypt" individual location information to protect users from being identified.

Existing studies have proposed several potential solutions for protecting geoprivacy. The simplest one is to aggregate fine-resolution data to upper-level scales. It indeed preserves user privacy but also reduces the spatial resolution of data [103]. The other commonly used method is grouping and mixing the geographic data (e.g., trajectory points) from k different users into k different regions and then generate k-anonymized location information [104]. Such a method may hide the spatial information of the input data and neglect temporal and semantic attributes. Another one is geomasking, which blurs users' locations by perturbation and adding noises so that the location information can be protected with spatial patterns preserved [105]. In addition, Rao [106] proposed a deep learning method using long short-term memory (LSTM) to generate a privacy-preserving synthetic trajectory that preserves the essential spatio-temporal attributes of the original trajectory [107]. All these studies may enhance the privacy protection of location-related information.

## 3.2  Lack of Direct Policy Implications

A major critique toward the spatial network literature is that the zeal to reveal spatial structures often does not lead to actual policy changes. Not many cities have changed their administrative boundaries according to emerged borders from spatial networks. One reason why these insights are only on the paper is due to the dynamic and uncertain nature of spatial structures derived from multiple sources [108, 109]. These boundaries are sensitive to the types of data collected and cannot represent the whole population. Spatial networks are also not the only way to delineate neighborhoods. It may have competing narratives with projects like Bostonography, which crowdsourced mental maps from users to represent conceptual neighborhoods [110]. Even if we take one of the derived spatial structure as the ground truth, very few papers went a step further to suggest a clear pathway for planners to act according to the local conditions or provide normative discussions on the results [117]. If we have spatial segregation in people's activity space, should we conform to such structure or optimize it to a healthier balance? If we know a city is at the margin of the urban hierarchy, how can we help the city move up the ladder or achieve a mutually beneficial state?"

   We outlined three potentials to answer the questions above. One of the missing pieces in the existing literature is the evolution of urban network structures and the associated socioeconomic phenomenon. Hidalgo [112] and Hausmann [113]'s economic complexity research showed an excellent example of linking international trade network insights into concrete economic development suggestions for developing countries. Planning literature lacks the equivalent depth of network knowledge on the city level to direct local economic policies, though Park [114]'s work on labor flows and geo-industrial clusters has started to tackle this challenge. Another way to connect network insights to policies is to evaluate the impact of policies that change network links causally, or vice versa, the effect of networks at implementing policies. Cao [115] and Liu [116] assessed the effect of the high-speed rail network on city-to-city travel time and air traffic distribution. Such research can help transportation planners to quantify the cost and benefits of actions that increase connectivity. Andris [117] and Goetz [118] also pointed out how policies are often applied to the regional level, while followed at the social networks level, such as the successes and failures in COVID-19 interventions. Tracing policy implementation through social networks may inform barriers or opportunities for planning actions to take place. Lastly, a more recent piece from Shelton [120] exemplified the power of contextualizing the spatial network structures of the inner-city to challenge the administrative boundaries. Shelton [120] made a compelling case for the City of Atlanta to reconsider the arrangements of Neighborhood Planning Units (a political legacy for the neighborhoods to rally and organize for their interests in urban planning) by tracing the historical evolution of neighborhoods and comparing it to the borders derived from big data. Validating the spatial networks insights with multiple sources and grounding them in contexts can further move policies forward.

### 3.3  Lack of Civic, Communicative, and Collaborative Engagement

When spatial network analysis first became popular, there were many excitements in the planning field to see its applications in supporting communicative and collaborative planning theories [3, 121]. However, as we observed in the recent empirical studies, very few consult or engage the information providers on interpreting and explaining the implications. The big geospatial data in spatial network research may produce an illusion that we have a representative picture of the whole population. What if there is a gap between how people conceive their activity spaces and what they show on their mobility trajectories? Furthermore, most of the network insights also do not directly serve the interests of local communities or non-profit organizations. While we often applaud the positive impacts of connectivity on social and economic welfare, we should not forget the duality of network edges: when an access point is not available, an edge, such as a highway, can also negatively affect the surrounding neighborhoods. The preferential attachment mechanism (i.e., rich gets richer) in scale-free networks also deepens inequality in connectivity distribution. How can communities use spatial network analysis to communicate for their rights and counter the privileged network discourse? How can we have more "human-in-the-loop" interpretations of spatial network insights?

The volunteered geographic information (VGI) literature in geography may provide some interactive models for planners to engage citizens in spatial network analysis. Participatory mapping and crowdsourcing can potentially be extended to spatial network information, though their validity can be difficult to confirm. A combination of recruited volunteers with GPS trackers and follow-up interviews can also provide valuable context to explain the motivations of emerging patterns [122].

Also, to generate counter-narratives, we need to think creatively about what constitutes alternative spatial network data. For instance, Andris [123] collaborated with the NGO *Big Brothers and Big Sisters of America* to conceptualize mentorship pairs in city as spatial networks and evaluated the impact of the mentoring program at bridging spatial gaps between places which will not be connected otherwise through commutes or demographic groups.

### 3.4 Difficulty to automate visualizations and integrate with GIS

Geovisualization and maps are essential in geographic information representation and GISystems. A precise visualization helps illustrate data information intuitively and vividly to the audience to better understand the story and turn it into knowledge. Though researchers have integrated various spatial network data into GISystems [124], visualizing spatial networks and interactions on maps still face conceptual and methodological challenges (see [125] for a comprehensive review). For example, traditional flow maps do not display well with large-scale non-planar networks as they can be too dense to display on a 2D space and thus result in edge cluttering. In addition, nodes and edges in spatial networks take on multiple attributes (e.g., density, direction, divisions, and hierarchies of flows and the attributes of the destination nodes), which requires more aesthetic support than just size and color in GIS.

   Researchers have proposed several methods to address these challenges. A set of algorithms have been proposed for visual simplification to reduce cluttering based on aggregation, such as automating thresholds to filter data [126]; grouping points that are close by or have similar connectivity into one node (i.e., graph partitioning [127] and spatial clustering [128, 129]); bundling and summarizing edges that are going to the same directions (i.e., edge bundling [130]); and algebraic multigrid [131]). Studies have also explored how to represent more attributes of networks in visualization, such as direction [132] and temporal or step-wise changes of flows (i.e., alluvial diagram [133]). Several powerful web-based visualization tools and packages were developed for spatial network visualization, such as deck.gl [134] and flowmap.blue[1]. Compared with traditional static GIS maps, these web-based interfaces can dynamically visualize more network attributes, such as the flow directions, the contrast of in and out-degree for nodes, and all connections to one node. All these studies may benefit spatial network-related geovisualization.

## 4 Conclusion: Envisioning a Collaborative Planning Model with Spatial Networks

Planning is a process for setting goals, identifying and assessing options, and developing strategies for achieving desired goals [135]. Collaborative planning theory further elaborates on the planning process to be an "interactive, communicative activity," [136, p.183] that engages diverse urban stakeholders [137]. Given the wide range of spatial networks applications, we believe that it can inform collaborations across agents and regions that enrich both the rationality and humanity of the outcomes. We proposed a framework in which spatial networks contextualize collaborative planning theory in practices and resolve some of the challenges we mentioned above (see Figure 1). In this idealized framework, private companies, governments, citizens, and researchers can all contribute spatial network data as a public good and participate in the generation of planning knowledge. The process will be facilitated by collaborations between various urban stakeholders that augments the interpretations of spatial network insights. In the end, planners can use network data and methods to convert planning knowledge into actions by evaluating planning alternatives, establishing expectations of planning impacts, or providing normative evidence to support activism. To further illustrate how different planning stakeholders can use spatial network analysis for decision-making or activism under collaborative planning framework, we generated a table of examples for each stakeholder (see Table 1).

---

[1] https://flowmap.blue/

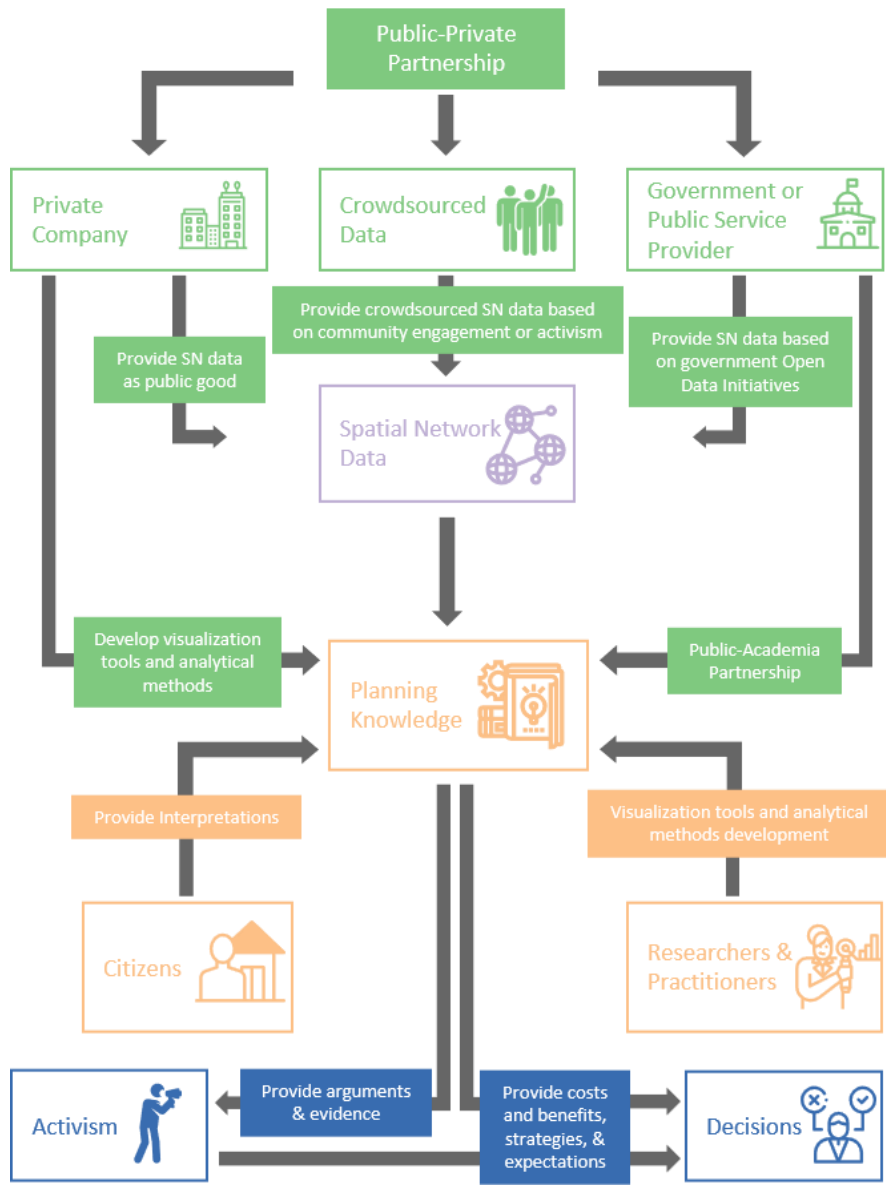

**Fig. 1** A collaborative planning framework with spatial networks

| Spatial Network Data | Methods | Example Actions |
|---|---|---|
| *Regional Commission* | | |
| Population density of cities (Nodes) and multiple urban networks through mobility, employment history, capital investments, and social media (Edges) | Apply community detection to delineate boundaries for megaregions and validate the boundaries with multiple types of networks | Propose urban clusters that can be developed into megaregions and suggest regional planning of infrastructure that can support such urban system |
| *Public Administration* | | |
| Real-time mobility tracking data with LBSN services during COVID-19 | Use community detections to capture urban clusters that have high risks of inter-city disease transmission based on connectivity | Revisit and update a dynamic administrative boundary for urban management and public health control by establishing inter-regions councils |
| *Tourism and Cultural Department* | | |
| Crowdsourced tourism stop points (Nodes) and routes (Edges) from travel agencies and locals | Convert each tourism route into a chained trip, identify places that are most popular and most likely to be a transfer stop, and cluster the trips based on different themes | Develop an automated system to recommend customized trips based on users' preferences |
| *Economic Planners* | | |
| Economic network constructed through company branches (Nodes) and firm collaboration records (Edges) | Calculate the probability of growing a industry based on connectivity to the industrial profiles from other cities | Strategize what industries to invest and grow not only based on a city's endogenous resources, but also based on the relative advantages of a city in urban networks |
| *Transportation Planners* | | |
| Fragmented bike lane network shapefile (Nodes and Edges) | Develop optimization algorithms to connect the fragmented bike lanes and evaluate the costs and benefits | Strategize where to build the next bike lane |
| *Urban Designers and Modelers* | | |
| Work-home commutes data (Nodes and Edges); Social Media Check-in (Nodes) | Approximate the density of visits and volumes of traffic flows in and out of a target area | Calibrate the model of passenger flow at a design site to simulate the effect of the proposed plan |
| *Housing and Community Planners* | | |
| Survey of people's active connections (Edges) within and outside of the current living communities (Nodes) | Construct a residential mobility index based on the neighborhood characteristics and people's social networks | Evaluate the effect of mixed-income public housing project on people's social capitals and job access |
| *Activists* | | |
| Collect survey of teachers' work-home locations (Nodes) and commute routes (Edges) | Visualize the information as a spatial network to show how far and scattered teachers currently live | Convince the local community to approve affordable housing for teachers so that they can live nearby the schools |
| *NGO and Social Enterprise* | | |
| Locations of coffee farms (Nodes) and their production relationships (Edges) | Interview to reveal collaborative and competitive relationships between local coffee farms and identify critical ties | Form local coffee producers co-op networks that improve economic and environmental resilience for individuals |

| Business owners and investors | | |
|---|---|---|
| Place-based or POI-based visit data (from social media) or LBSN services | Evaluate the sociodemographic profiles of people that frequent the locations and their matches to the business | Select optimal locations to open new business |

**Table 1** Example applications of spatial networks for planning stakeholders.

In conclusion, we reported research that applied spatial networks data and methods in the planning context. We found four common themes, including revealing spatial structures, optimizing urban infrastructure, correlating network structure with economic development, social capitals, and residential mobility, and monitoring public health. We also discussed challenges of data openness and privacy, unclear policy implications, lack of civic engagement, and difficulty to integrate with GIS that impeded planners' further adoption of spatial network analysis in daily routines. However, we are optimistic that spatial networks can be the backbone of a collaborative planning framework. The hypothetical examples we provided from various urban stakeholders' perspectives show spatial networks' flexibility to facilitate efficiency, responsiveness, and inclusion in planning practices. Future research should address the applications in environmental, ecological, and energy networks that are not covered in this study.